\documentclass[aps,pra,twocolumn,groupedaddress,showpacs]{revtex4}

\usepackage{graphicx}
\usepackage{graphicx}
\usepackage{amssymb}
\usepackage{amsmath}
\usepackage{color}
\newcommand{\be}{\begin{equation}}
\newcommand{\ee}{\end{equation}}
 
 \definecolor{BrickRed}{cmyk}{0,0.89,0.94,0.28}
\definecolor{MidnightBlue}{cmyk}{0.98,0.13,0,0.43}
\definecolor{DarkGreen}{rgb}{0,0.7,0.1}

\begin{document}

\title{Probing the non-Planckian spectrum of thermal radiation in a micron-sized cavity  with a spin-polarized  atomic beam}


\author{Giuseppe Bimonte}

\affiliation{ Dipartimento di Fisica E. Pancini, Universit\`{a} di
Napoli Federico II, Complesso Universitario
di Monte S. Angelo,  Via Cintia, I-80126 Napoli, Italy}
\affiliation{INFN Sezione di Napoli, I-80126 Napoli, Italy}

\email{giuseppe.bimonte@na.infn.it}

\begin{abstract}

It is commonly thought that thermal  photons with transverse electric polarization cannot exist in a planar metallic cavity whose size $a$ is smaller than the thermal wavelength $\lambda_T$,  due to absence  of modes with  $\lambda < 2a$.   Computations based on a realistic model of the mirrors contradict this expectation, and show that a micron-sized metallic cavity is  filled with non-resonant radiation having transverse electric polarization,  following a non-Planckian spectrum, whose average density at room temperature  is  orders of magnitudes larger than that of a black-body.  We show that  the spectrum of this radiation  can be measured  by observing the transition rates  between hyperfine ground-state sub-levels $1S_{1/2}(F,m_F) \rightarrow 1S_{1/2}(F',m'_F)$  of D atoms passing in the gap between  the mirrors  of a   Au cavity.  Such a measurement would also shed light 
on a  puzzle in the field of dispersion forces, regarding the sign and magnitude of the thermal   Casimir force. Recent experiments with Au surfaces led to contradictory results, whose interpretation  is  much controversial.

\end{abstract}

\pacs{12.20.-m, 
03.70.+k
,42.25.Fx 
}

\maketitle

\section{Introduction}
\label{sec:intro}

The  study of  the spectrum of the electromagnetic (em) fluctuations in a cavity in thermal equilibrium is at the roots of quantum mechanics. 
The fundamental importance of this problem was  recognized after Kirchhoff  showed
that the energy density of thermal radiation is   a universal function of the frequency and of the temperature,  independent of the material properties of the walls. This discovery stimulated intense efforts to  work out theoretically the analytical expression of this universal function, which was finally  found by M. Planck \cite{Planck}. 

The famous theorems of Kirchhoff are valid in a large box, and at distances from the walls much larger than the wavelength. This observation naturally  leads one to wonder how the Planck spectrum gets modified in a small cavity, and/or in proximity of its walls. The study of this important problem was initiated by Rytov \cite{rytov}, the founder of the modern theory of em fluctuations, who computed the thermal field in proximity of a dielectric slab.  Since then, Rytov's theory  turned into  a vast field of research that goes by the name of fluctuational electrodynamics, with many diverse applications extending from heat radiation to heat transfer, as well as to  Casimir and van der Waals forces both in  and  out equilibrium    \cite{kruger,mehran}. 

The fundamental result of fluctuational electrodynamics is the following formula for the  correlator of the em field in a vacuum region of space, surrounded by any number of dielectric bodies at temperature $T$, which can be derived by linear-response theory \cite{agarwal}: 
\begin{eqnarray}
& & \langle {\hat E}_{\alpha}({\bf r},t) {\hat E}_{\beta}({\bf r'},t') \rangle_{\rm sym}= \hbar \int_{-\infty}^{\infty}\!\! \frac{d \omega}{2 \pi} \coth\left(\frac{\hbar \omega}{2 k_B T}\right)\nonumber \\
& & \;\times\; {\rm Im}[{\cal E}_{\alpha \beta}({\bf r},{\bf r'},\omega)] \;e^{-i \omega(t-t')} \;,\label{corr}
\end{eqnarray}  
where   ${\cal E}_{\alpha \beta}({\bf r},{\bf r'},\omega)$ is the Fourier transform of the classical Green function of the electric field, obtained by solving the {\it macroscopic} Maxwell Equations in the background of the   bodies. 
The  correlator of the magnetic field  has an analogous form, in terms of the  magnetic Green function ${\cal H}_{\alpha \beta}({\bf r},{\bf r'},\omega)$. 
The subscript sym in the average symbols denotes symmetrized products of field operators \footnote{The correlators for different orderings of the operators can be expressed  in terms of the symmetrized one \cite{agarwal}}.
We stress that Eq.(\ref{corr}) describes both {\it zero-point} (i.e. quantum) and {\it thermal} fluctuation of the em field. The validity of Eq. (\ref{corr}) is only
subjected to the condition that the length scales of the relevant fluctuations should be large compared to atomic distances, in  such a way that  the em  material properties can be described by   {\it macroscopic} response functions.

A wealth of quantum em phenomena can be studied using the correlator in Eq. (\ref{corr}). For example, starting from the free-space Green function, it is easy to derive from Eq. (\ref{corr}) Planck's law, and to compute the coefficient of spontaneous emission of an atom in vacuum \cite{wylie}. The most spectacular applications of Eq. (\ref{corr}) are however in bounded geometries. Consider for example a  planar cavity consisting of two mirrors separated by a vacuum gap. By evaluating the difference among the averages of the Maxwell stress tensor at a point inside the gap and a point just outside the cavity, one can obtain the celebrated  Lifshitz formula  \cite{lifs} for the Casimir pressure between two dielectric slabs at finite temperature, which generalized to real media the famous formula of Casimir for ideal mirrors at zero temperature \cite{Casimir48}.  Other classic cavity-QED problems that can studied with the help of Eq. (\ref{corr})  include the Casimir-Polder interaction of an atom with one or more material walls \cite{haroche,buh}, and  suppression of spontaneous decay of an atom in a cavity \cite{wylie,kleppner,jhe}. Applications of Eq. (\ref{corr}) are endless, being limited only by one's ability to work out the classical Green functions in more complex geometries.

In this paper we use fluctuational electrodynamics to study the thermal radiation existing in a micron-sized planar metallic cavity, and we describe the scheme of an experiment with spin-polarized D atoms to observe its spectrum. 

On general grounds, the spectrum of the thermal radiation of a narrow cavity is expected to deviate strongly from the Planckian form. Our analysis shows  that the properties of the obtained radiation depend dramatically on whether the mirrors are modeled as lossy or rather as dissipation-less. When dissipation is neglected,  the radiation is mostly constituted   (for moderately high temperatures) by em fields with transverse magnetic (TM) polarization, while barely any radiation with transverse electric (TE) polarization is present. This is consistent with one's expectation that no photons with TE polarization can propagate in a narrow cavity whose width $a$ is smaller than thermal length $\lambda_T=\hbar c/k_B T$ ($\lambda_T=6.7\;\mu$m for $T=300$ K). Surprisingly, the features of the radiation change drastically when dissipation of the mirrors is taken into account. Computations show that  a lossy cavity is indeed  filled with non-resonant TE-polarized  radiation, having a broad non-Planckian spectrum, whose density is orders of magnitude larger than that of a large black-body (BB) cavity at the same temperature.  Closer inspection shows that this intense radiation is mainly constituted by thermally excited magnetic fields. 

The existence of thermal fluctuations of the magnetic field  in proximity of a {\it single} metallic surface has been pointed out prior to the present work \cite{carsten1,carsten3},  in  studies of cold atoms trapped in magnetic quadrupole traps. Since these traps can only hold atoms whose magnetic moment is aligned with the magnetic field in the trap center,   spin flips caused by the magnetic noise may lead to   losses of atoms from the trap.   Measurements of   the escape rate from traps approaching the surfaces of different metals  \cite{cornell}, down to a distance of  5 micron,  are in qualitative agreement with theoretical calculations,  based on the lossy Drude model   \cite{carsten1,carsten3}. 

The   consequences of  magnetic noise  for the thermal energy of  a cavity,  which is the object of the present work, have not been explicitly worked out before. There are very good reasons to consider this problem as being worth of  detailed   theoretical and experimental investigations. Observing the spectrum   of the thermal radiation existing in a micron-sized cavity would indeed  provide valuable insights to help resolving a long-lasting puzzle in Casimir physics, concerning the magnitude and the sign of the thermal correction to the Casimir force between two metallic bodies. Computations (see Sec. III) show that the thermal magnetic fields predicted by the lossy model of the cavity lead to  a tiny   {\it repulsive} force between the mirrors, which  coincides with the thermal correction to the Casimir force predicted by Lifshitz theory \cite{sernelius}. The problem is that   two series of precision experiments  with Au test bodies at sub-micron separations, carried out by two different groups,  \cite{decca1,decca2,decca3,decca4,chang,bani1,bani2}  found no evidence of the thermal force. The  data of these  experiments  are inconsistent  (up to 99 \% c.l.) with the theoretical analysis of \cite{sernelius}. Surprisingly, the data  are instead consistent  (up to 90 \% c.l.) with Lifshitz theory, provided that  conduction electrons are  modeled  as a dissipation-less plasma. When the latter model is used to compute the Casimir force,  the obtained thermal  correction is {\it attractive} for all separations, and for $a \gg \lambda_p$   it is undistinguishable from that for a perfect conductor (more details can be found in the book \cite{book2}). To explain the findings of these experiments, it  has been  conjectured \cite{sernelius2}  that when two conductors are brought into close proximity,   saturation effects  may lead to  suppression of thermal fluctuations of the em field, which might explain why the repulsive thermal  Casimir force  predicted by the Drude model has not been seen in the experiments. 
Agreement with the Drude model has been reported in a single torsion-balance experiment \cite{lamorth}, which probed the Casimir force in the range from 700 nm up to the large separation of 7.3 $\mu$m.   
The  interpretation of the latter experiment is, however, obscured by the fact that the Casimir force  was not measured directly, but rather  estimated indirectly after subtracting from the data  a much larger force (up to one order of magnitude), supposedly originating from electrostatic patches, by a fit procedure based on a phenomenological model of the unknown electrostatic force.   We point out that an  attempt to measure the thermal Casimir force in a plane-parallel Al setup at separations larger than three micron was  reported earlier in \cite{antonini}, but observation of the Casimir force turned out impossible due to   the presence of large unexplained forces, presumably of electrostatic origin.  Finally, we note that a new experiment to probe the gradient of the Casimir force between two parallel plates at large separations is presently  under way  \cite{rednik}. 

These considerations show that it would be of great interest to {\it directly} observe the spectrum of the thermal radiation of a metallic cavity, to make sure that the Drude model provides the correct description. We show that this could be done using a spin-polarized beam of D atoms passing in the gap between the mirrors of a micron-sized cavity.

The plan of the paper is the following: in Sec. II we compute the thermal radiation existing in a micron-sized  metallic plane-parallel cavity,  and work out its non-Planckian spectrum. In Sec III we discuss the connection between the thermal radiation of the cavity and  the thermal Casimir force.  In Sec. IV we show that the spectrum of this radiation can be measured by observing the transition rates among hyperfine ground-state sub-levels  of  spin-polarized  D atoms passing between the mirrors. Finally in Sec. IV we present our conclusions.

\section{Thermal radiation in a plane-parallel cavity}

According to intuition, very few thermal photons with transverse electric (TE) polarization can exist in the cavity  if its width $a$ is smaller than the thermal wavelength $\lambda_T$   ,  since  their wavelengths $\lambda \simeq \lambda_T$ are just too long to fit in the narrow gap between the mirrors. The absence of photon states with TE polarization for $\lambda < 2 a$ is indeed  the main reason for the suppression of spontaneous emission by excited Cs atoms with maximal angular momentum normal to the mirrors reported in \cite{jhe}.   
These considerations lead  one to  expect that  for $a \ll \lambda_T$ there should be very little thermal energy in the cavity,  in the form of em fields with TE polarization. Surprisingly, this expectation appears to be incorrect. Let us see how this comes about.  

\begin{figure}
\includegraphics [width=.9\columnwidth]{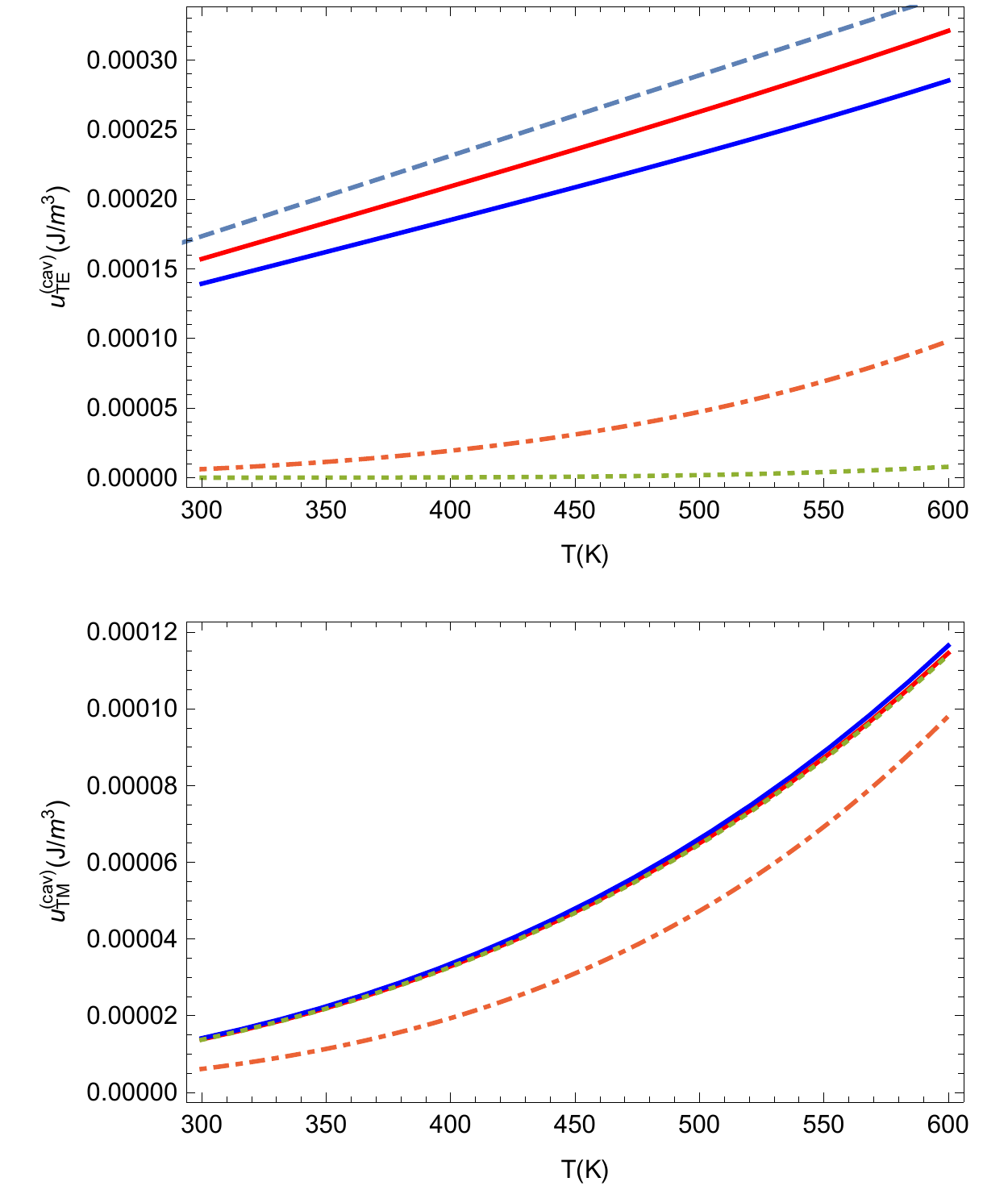}
\caption{\label{energyT}  Energy density of  TE (upper panel) and  TM (lower panel) modes  at   the center of a 2 $\mu$m cavity, versus temperature (in K).   In both panels, the upper and lower solid lines are for two lossy mirrors made of Au and Pt, respectively. The lower dotted line is for a Au cavity with dissipation neglected. The dot-dashed curve shows the energy density of a black-body. The upper dashed line in the upper panel  shows the universal energy density in Eq. (\ref{univ}). }
\end{figure}

To be definite, let us set a cartesian coordinate system $(x,y,z)$ such that the $z$ coordinate spans the axis normal to the mirrors, which have coordinates $z=0$ and $z=a$, respectively.  The (unrenormalized) energy density $u_{\rm unr}^{\rm (cav)}(z)$ at a point $P$ of coordinate $z$  inside the cavity  is equal to the average of the time component $T_{00}(z)$ of the Maxwell stress-energy tensor at $P$:
\be
u_{\rm unr}^{\rm (cav)}(z) =  \sum_{\alpha}\frac{ \langle {\hat E}_{\alpha}^2 ({\bf r}, t) \rangle+ \langle {\hat B}_{\alpha}^2 ({\bf r}, t) \rangle}{8 \pi} 
\ee
Using the general formula for the correlators Eq. (\ref{corr}) we obtain:
$$
u_{\rm unr}^{\rm (cav)}(z) =   \frac{\hbar}{4 \pi} \int_{0}^{\infty}\!\! \frac{d \omega}{2 \pi}  \coth\left(\!\frac{\hbar \omega}{2 k_B T}\!\right) \! 
$$
\be
\times \sum_\alpha \left ({\rm Im}[{\cal E}^{\rm (cav)}_{\alpha \alpha}({\bf r},{\bf r},\omega)]+
{\rm Im}[{\cal H}^{\rm (cav)}_{\alpha \alpha}({\bf r},{\bf r},\omega)] \right)\;\label{bareen}
\ee 
The explicit expression of the electric and magnetic Green functions of the cavity, ${\cal E}^{\rm (cav)}_{\alpha \alpha}({\bf r},{\bf r},\omega)$ and ${\cal H}^{\rm (cav)}_{\alpha \alpha}({\bf r},{\bf r},\omega)$ can be found in the Appendix. We note that in writing Eq.  (\ref{bareen}) we took advantage of the fact that the imaginary parts of the Green functions are odd functions of the frequency $\omega$, to express the energy as an integral over positive frequencies only.
Equation (\ref{bareen}) is formally divergent, but it can be easily renormalized by  decomposing the hyperbolic cotangent (times $\hbar/2$) as
\be
\frac{\hbar}{2} \coth\left(\frac{\hbar \omega}{2 k_B T}\right)={\rm sgn} (\omega) \left(\frac{\hbar}{2}+ \frac{\hbar}{\exp(\hbar |\omega|/k_B T)-1} \right)\,
\ee
The $\hbar/2$ term on the r.h.s. of the above Equation is interpreted as describing the contribution of quantum zero-point fluctuations of the em field, while the Bose-Einstein term represents the contribution of thermally excited em fields.  After the above  identity is substituted  into Eq. (\ref{bareen}), one further notes that the  representation  of the cavity Green function as the sum of the  free-space  Green function   plus a scattering contribution   (see Eqs. (\ref{Greensplit}) and (\ref{GreensplitH})) allows to decompose the unrenormalized energy density  $u_{\rm unr}^{\rm (cav)}(z)$ as the sum of four terms:
\be
u_{\rm unr}^{\rm (cav)}(z)=u^{(0)}_{\rm z.p.}+u^{(\rm sc)}_{\rm z.p.}(z)+u_{\rm BB}+u^{(\rm sc)}(z)\;,\label{ubare}
\ee
where
\begin{eqnarray}
&&u^{(0)}_{\rm z.p.}= \frac{\hbar }{\pi c^3} \int_{0}^{\infty}\!\! \frac{d \omega}{2 \pi} \omega^3 \;,\nonumber \\ 
&&u^{(\rm sc)}_{\rm z.p.}(z)=\hbar \int_{0}^{\rm \infty} \frac{d \omega}{2 \pi}   \int \frac{d^2 {\bf k}_{\perp}}{(2 \pi)^2} \;  g(\omega,{\bf k}_{\perp})\;, \nonumber \\
&&u_{\rm BB}= \frac{2 \hbar }{\pi c^3} \int_{0}^{\infty}\!\! \frac{d \omega}{2 \pi}  \frac{ \omega^3}{\exp(\hbar \omega/k_B T)-1} \;,
\end{eqnarray}
and
\be
u^{(\rm sc)}(z)=	\! \int_{0}^{\rm \infty} \!\!\frac{d \omega}{2 \pi}  \frac{2\, \hbar}{\exp(\hbar \omega/k_B T)-1} \!\! \int \!\!\frac{d^2 {\bf k}_{\perp}}{(2 \pi)^2}   g(\omega,\!{\bf k}_{\perp}\!).
\ee
The function $g(\omega,{\bf k}_{\perp};z)$ is:
$$
g(\omega,{\bf k}_{\perp};z)=k_{\perp}^2 \sum_{\alpha={\rm s,p}} {\rm Re}\left[\frac{1}{k_z}\left( \frac{R_{\rm \alpha}^{(1)} }{ {\cal A}_{\rm \alpha}}e^{2 i k_z z} \right.  \right.
$$
\be
\left. \left. +  \frac{R_{\rm \alpha}^{(2)} }{{\cal A}_{\rm \alpha}}e^{2 i k_z (a-z)} \right)+ \frac{2\,\omega^2}{c^2 k_{\perp}^2}  \frac{R_{\rm \alpha}^{(1)}  R_{\rm \alpha}^{(2)} }{ {\cal A}_{\rm \alpha}}e^{2 i k_z a}  \right]
\ee
Of the four terms  on the r.h.s. of Eq. (\ref{ubare}), only the first one $u^{(0)}_{\rm z.p.}$ is divergent, while the remaining three terms are finite. The divergent term $u^{(0)}_{\rm z.p.}$  represents the unobservable  energy of zero-point fluctuations in free-space. After disregarding this term,  one gets the following well-defined expression for the {\it total} energy density in the gap:
\be
u_{\rm ren}^{\rm (cav)}(z)= u^{(\rm sc)}_{\rm z.p.}(z)+u_{\rm BB}+u^{(\rm sc)}(z)\;.
\ee  
The first  term $u^{(\rm sc)}_{\rm z.p.}(z)$ on the r.h.s. of this Equation  is interpreted as  representing the {\it shift} of the zero-point energy due to scattering of virtual photons by the mirrors, while the second term $u_{\rm BB}$ coincides with Planck's formula for the energy density of a large black-body cavity. Finally, the third term  $u^{(\rm sc)}(z)$ provides the correction to Planck's formula arising  from (multiple) scatterings of thermally excited photons by the mirrors. According to the physical interpretation  of the three terms, we define the {\it thermal} energy density of the cavity  $u^{(\rm cav)}(z)$ by the formula:
\be
u^{(\rm cav)}(z)=u_{\rm BB}+u^{(\rm sc)}(z)\;.
\ee
It is important to observe that the function $g(\omega,{\bf k}_{\perp};z)$  involves the Fresnel reflection coefficients $R_{\rm \alpha}^{(k)}$ of the mirrors, which in turn depend on their (complex) permittivity $\epsilon(\omega)$.  One notes at this point that the wavelengths $\lambda \simeq \lambda_T$ of thermal photons  belong to the infrared region, where metals display little dissipation. This consideration suggests that for a realistic modeling  of the cavity, which takes into account the  finite skin depth of em fields in real metals,  it is sufficient to model  the mirrors by the {\it dissipation-less} plasma model of infrared optics, according to which:
\be
\epsilon(\omega)=1-\omega_{\rm p}^2/\omega^2+\epsilon_{\rm core}(\omega)\;,\label{plasma}
\ee 
where $\omega_{\rm p}$ is the plasma frequency and $\epsilon_{\rm core}(\omega)$ is the contribution of bound electrons.  When  the dissipation-less model in Eq. (\ref{plasma}) is used, the obtained spectrum actually confirms the initial expectation, showing that in the cavity there are barely any  thermal photons with TE polarization. For example, at the center of a 2 $\mu$m Au cavity at 300 K the energy density ${u}^{(\rm cav)}_{\rm TE}\vert_{\rm pl}$ associated 
with TE photons  turns out to be seven hundred times smaller than the energy density ${u}_{\rm BB}$ of a (large) black-body (BB) cavity at the same temperature (see the dotted line in the upper panel of Fig.\ref{energyT}).   

\begin{figure}
\includegraphics [width=.9\columnwidth]{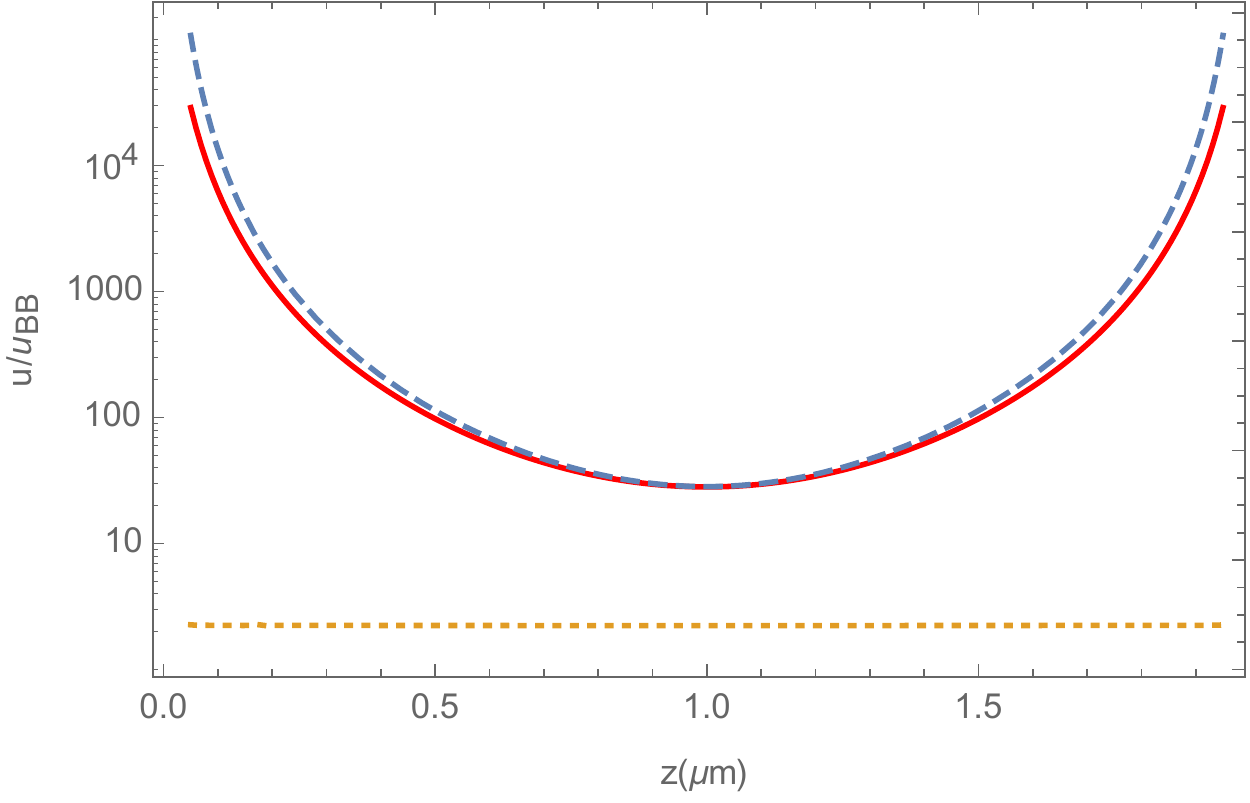}
\caption{\label{ucavvsz}  Normalized   energy density  $u^{(\rm cav)}(z)/u_{\rm BB}$ inside a 2 $\mu$m Au cavity at room temperature, versus the position  $z$ (in micron). The solid and the dotted lines correspond, respectively, to lossy and to dissipation-less mirrors. 
The upper dashed line is a plot of the universal energy density in Eq. (\ref{univ}). }
\end{figure}

Real metals however do display dissipation at all frequencies,  and one thus considers that a more accurate picture of the spectrum of the cavity would be obtained   by using the Drude dielectric function for the conduction electrons:  
\be
\epsilon (\omega)=1-\omega_{\rm p}^2/\omega(\omega+i \gamma) + \epsilon_{\rm core}(\omega)\;, \label{drude}
\ee 
where $\gamma$ is the (temperature-dependent) relaxation frequency.   The results are surprising, for one finds that dissipation does not engender, as one would expect,  just a small  correction to the energy density $u^{(\rm cav)}(z)$  obtained by using the plasma model, but instead leads to a huge increase of the energy density. This can be seen from Fig. \ref{ucavvsz}, where the energy $u^{(\rm cav)}(z)$, normalized by the BB energy $u_{\rm BB}$, is displayed for values of $z$ corresponding to points whose minimum distance from the walls if larger than 50 nm. The solid and the dotted lines correspond to inclusion and to neglect of dissipation, respectively.
As we see, the energy density for dissipation-less mirrors is about twice $u_{\rm BB}$, and is nearly constant for the considered values of $z$, while the energy density  of the lossy cavity is everywhere larger than $u_{\rm BB}$ by orders of magnitudes, and increases enormously as $z$ approaches the mirrors. We observe that both $u^{(\rm cav)}(z)$ and $u^{(\rm cav)}(z)\vert_{\rm pl}$ diverge on the surface of the mirrors.  One should recall however that the macroscopic theory loses its validity at points whose separation from the mirrors is not large compared to atomic distances.  It turns out that the  contribution  $u^{(\rm cav)}_{\rm TM}$   of em fields with transverse magnetic (TM) polarization  is unaffected by the inclusion of dissipation, $u^{(\rm cav)}_{\rm TM} \simeq u^{(\rm cav)}_{\rm TM}\vert_{\rm pl}$, and therefore the huge increase of the energy caused by dissipation is due to modes with TE polarization. This is explicitly shown  
by Fig.\ref{energyT}, where the energy densities of TE modes (upper panel) and TM modes (lower panel) at the center of the cavity are plotted as a function of the temperature $T$ (in K).
In both panels the upper red and the lower blue  solid lines correspond, respectively, to  lossy mirrors made of Au  or Pt, while the dotted lines correspond to neglect of dissipation in a Au cavity, and the dot-dashed lines show the BB energy density.  
The lower panel demonstrates that the energy of TM modes is practically independent of the conductivity of the plates. The upper panel shows that the energy of TE modes increases by many orders of magnitudes as one includes the effect of dissipation.   

\begin{figure}
\includegraphics [width=.9\columnwidth]{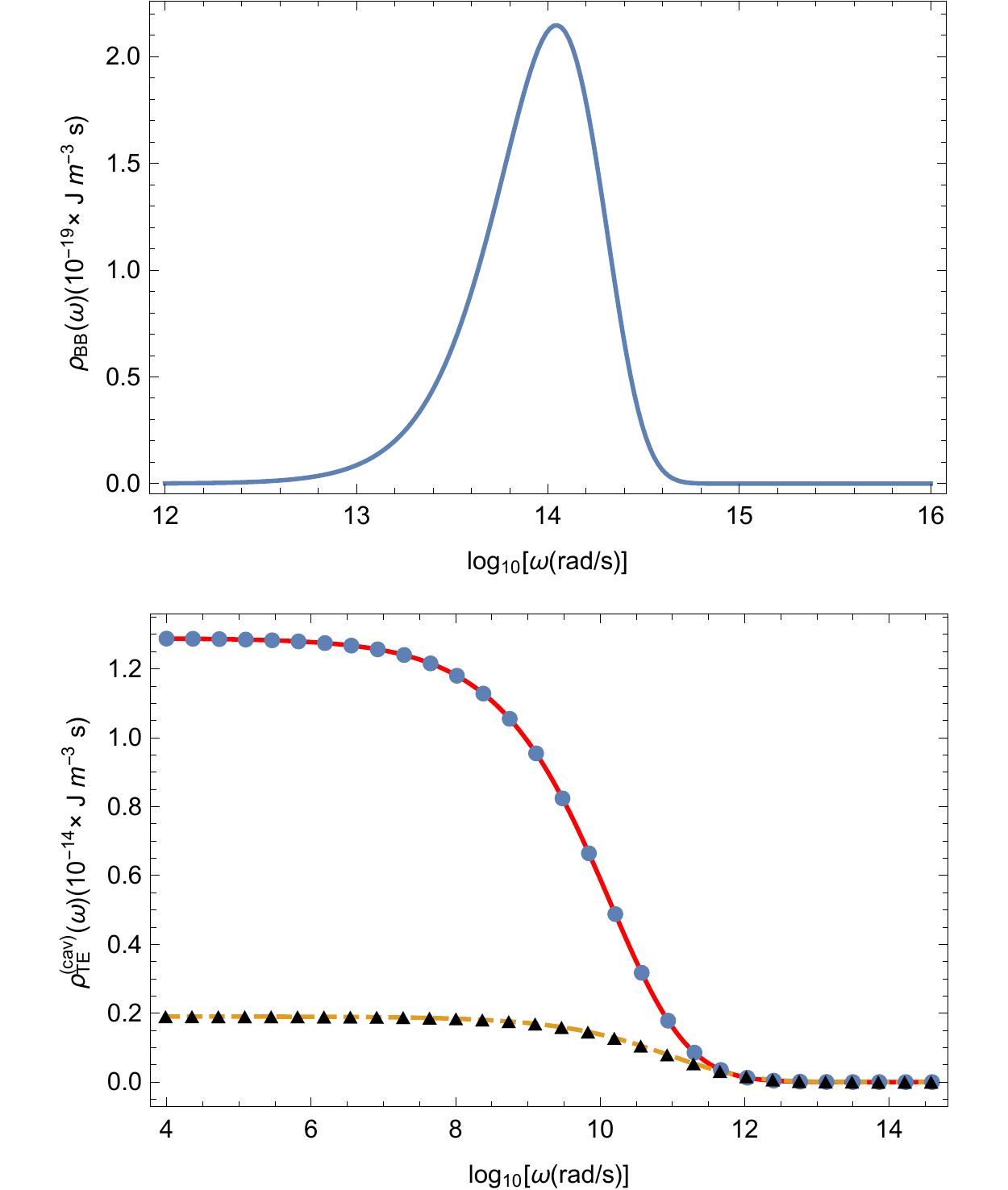}
\caption{\label{energy}   Spectrum of the room temperature energy density of TE modes  at the center of a 2 $\mu$m cavity (lower panel) made of Au (solid line) or of Pt (dotted line), versus angular frequency $\omega$.  The full circles and the triangles show the  spectrum of the magnetic energy. The mirrors are modeled as lossy conductors. The corresponding spectrum for loss-less mirrors is negligibly small and is not displayed in the Figure. For comparison, the upper panel shows the spectrum of a black body at the same temperature.}
\end{figure}

A better insight into the origin of the large  energy $u^{(\rm cav)}_{\rm TE}$ can be obtained by looking at its spectrum $\rho^{\rm (cav)}_{\rm TE} (\omega)$ \footnote{The angular-frequency spectrum  of the energy density (and of any other quantity) is defined such that ${u}^{\rm (cav)}= \int_0^{\infty} d\omega/2 \pi \; \rho^{\rm (cav)} (\omega)$.}, say at the centre of the cavity. Plots of $\rho^{\rm (cav)}_{\rm TE} (\omega)$ are  shown in  the lower panel of Fig.\ref{energy} for two lossy mirrors made of Au (solid line) or Pt (dotted line) (in the upper panel we show for comparison the Planck's spectrum).  The corresponding spectrum from dissipation-less mirrors is negligibly small, and is not displayed in the Figure.  The plot in Fig.\ref{energy}  shows  that   $u^{(\rm cav)}_{\rm TE}$ has a broad spectrum extending from zero-frequency up to a maximum frequency  $\tilde \omega \simeq  \omega_c^2/(4 \pi \sigma)$, where $\omega_c=c/2a$ is the characteristic cavity frequency and $\sigma$ is the conductivity. In micron-sized cavities $ \tilde{\omega} \ll \omega_T=k_B T/\hbar=3.9 \times 10^{13}\times(T$/300 K) rad/s. Indeed  for a Au cavity  $\tilde \omega\vert_{\rm Au}=6.4 \times 10^{9}\times (1 \mu{\rm m}/a)^2$ rad/s,   while for a Pt cavity $\tilde \omega\vert_{\rm Pt}=4.3 \times 10^{10}\times  (1 \mu{\rm m}/a)^2$ rad/s. It is interesting to note that the spectrum  of $\rho^{\rm (cav)}_{\rm TE} (\omega)$ has a tail extending up to frequencies as large as $\omega=$ (10$\div$100) $\tilde \omega$, which contribute significantly to the energy density $u^{\rm (cav)}_{\rm TE}$. The tail explains why the energies $u^{\rm (cav)}$ of Pt and Au are not much different (see Fig.\ref{energyT}), despite the factor of 6 difference in their spectral densities for low $\omega$ visible in Fig.\ref{energy}.   Inspection  of the magnitudes of the in-plane wave-vectors ${\bf k}_{\perp}$ contributing to $u^{\rm (cav)}_{\rm TE}$ reveals that  $u^{\rm (cav)}_{\rm TE}$  is associated with {\it evanescent} TE modes, consisting predominantly of {\it magnetic fields}. The latter property is demonstrated by the coincidence of the solid and dashed lines in Fig.\ref{energy} with the full circles and the triangles, respectively, which show the magnetic contribution to $\rho^{\rm (cav)}_{\rm TE} (\omega)$.
Fig.\ref{energy} shows that differently from Planck's law, the   spectrum $\rho^{\rm (cav)}_{\rm TE} (\omega)$   is {\it not} a universal function of the frequency and of the  temperature,  for it  depends  on the  conductivity of the mirrors (and also on their thickness). Quite remarkably, though, it can be shown that if the width $a$ of the cavity and the mirror thicknesses $w$ satisfy the conditions $\lambda_{\rm p} \ll a \ll \lambda_T$ and $w \gg \lambda_{\rm p}$, where $\lambda_{\rm p}=c/\omega_{\rm p}$ is the plasma length     ($\lambda_{\rm p}=22$ nm for Au),   at points $P$ in the gap whose (minimum) distance $d$  from the mirrors  satisfies the condition $d \gg \lambda_{\rm p}$, the energy density ${u}^{(\rm cav)}_{\rm TE}(P)$ of the TE modes    attains the {\it universal} value:
\be
{\tilde u}^{(\rm cav)}_{\rm TE}(z)=\frac{k_B T}{16 \pi a^3}\left[\zeta(3,z/a)+\zeta(3,1-z/a) \right]\;,\label{univ}
\ee  
where $\zeta(s,a)$ is generalized Riemann zeta function. This Equation indicates that the fluctuations  contributing to ${\tilde  u}^{(\rm cav)}_{\rm TE}$ are {\it classical},    and that the energy of the TE modes grows linearly with the temperature, differently from the $T^4$ behavior of the BB energy ${ u}_{\rm BB}$. A plot of ${\tilde u}^{(\rm cav)}_{\rm TE}(z)$ is shown by the upper dashed line of Fig. \ref{ucavvsz}, and by the (upper) dashed line in the upper panel of Fig. \ref{energyT}. 

Summarizing, we have shown that when the mirrors of a cavity are modeled as  lossy conductors, the cavity gets filled with thermal magnetic fluctuations  whose   energy density is larger by orders of magnitudes than that of a BB at the same temperature (at least for moderately high temperatures). The physical source of these magnetic fields can be traced back to the thermal electronic noise (Johnson-Nyquist noise) of a ohmic conductor at finite temperature. According to the fluctuation-dissipation theorem \cite{landau} the power spectrum of the Johnson-Nyquist noise is in fact proportional to the imaginary part of the permittivity $\epsilon(\omega)$,  which explains why the phenomena described here  become manifest only when the mirrors  are treated as dissipative.

\section{The puzzle of the thermal Casimir force}

The thermal spectrum    shown in Fig.\ref{energy}  has never been tested experimentally, and so it is legitimate to wonder if the theoretical framework which led to it is right. The question is not idle, because {\it indirect} evidence from recent Casimir experiments raises more than a doubt.  Casimir experiments   indeed  probe the {\it mechanical} effects of  thermal em fields on the mirrors. We said earlier that Lifshitz computed the Casimir pressure between two mirrors   by evaluating the average of the Maxwell stress tensor over the quantum and thermal fluctuations of the em field in the gap between the mirrors \cite{lifs}, and then subtracting its average at a point just outside the cavity. When Lifshitz formula is computed for two lossy mirrors at finite temperature \cite{sernelius} the obtained Casimir pressure $P_{\rm C}(a;T)$ includes automatically both the contribution $P^{(\rm zp)}_{\rm C}(a)$ of zero-point fluctuations and the contribution $\Delta P_{\rm C}(a;T)$  of thermal fluctuations. For separations $\lambda_{\rm p} \ll a \ll \lambda_T$,    $\Delta P_{\rm C}(a;T)$   is {\it repulsive} and has magnitude:
 \be
\Delta P_{\rm C}(a;T) \simeq \frac{k_B T}{8 \pi a^3} \zeta(3) \left [1- 6 \,\frac{\lambda_{\rm p}}{a}+ O\left( \frac{\lambda_{\rm p}}{a} \right)^2 \right]\,,\label{thforce}
\ee
where $\zeta(3)=1.202$ is Riemann zeta function. It can be verified that $\Delta P_{\rm C}(a;T)$  coincides with the Maxwell stress  associated with the thermal magnetic fields contributing to ${  u}^{(\rm cav)}_{\rm TE}$,  that we described  earlier.  The repulsive thermal force has been indeed interpreted as the effect of the magnetic coupling among the Johnson-Nyquist thermal currents  with the eddy currents induced by them  in the opposite plate \cite{BimonteJ,carsten}.   The existence of a repulsive thermal correction to the Casimir force between two lossy mirrors,  was noted for the first time in \cite{sernelius}, and gave rise to intense efforts to observe it. 
As we pointed out earlier two series of precise experiments \cite{decca1,decca2,decca3,decca4,chang,bani1,bani2}  found no evidence of the repulsive force in Eq. (\ref{thforce}) predicted by the lossy model of the mirrors, and were instead consistent with a loss-less model of the mirrors.

\begin{figure}
\includegraphics [width=.9\columnwidth]{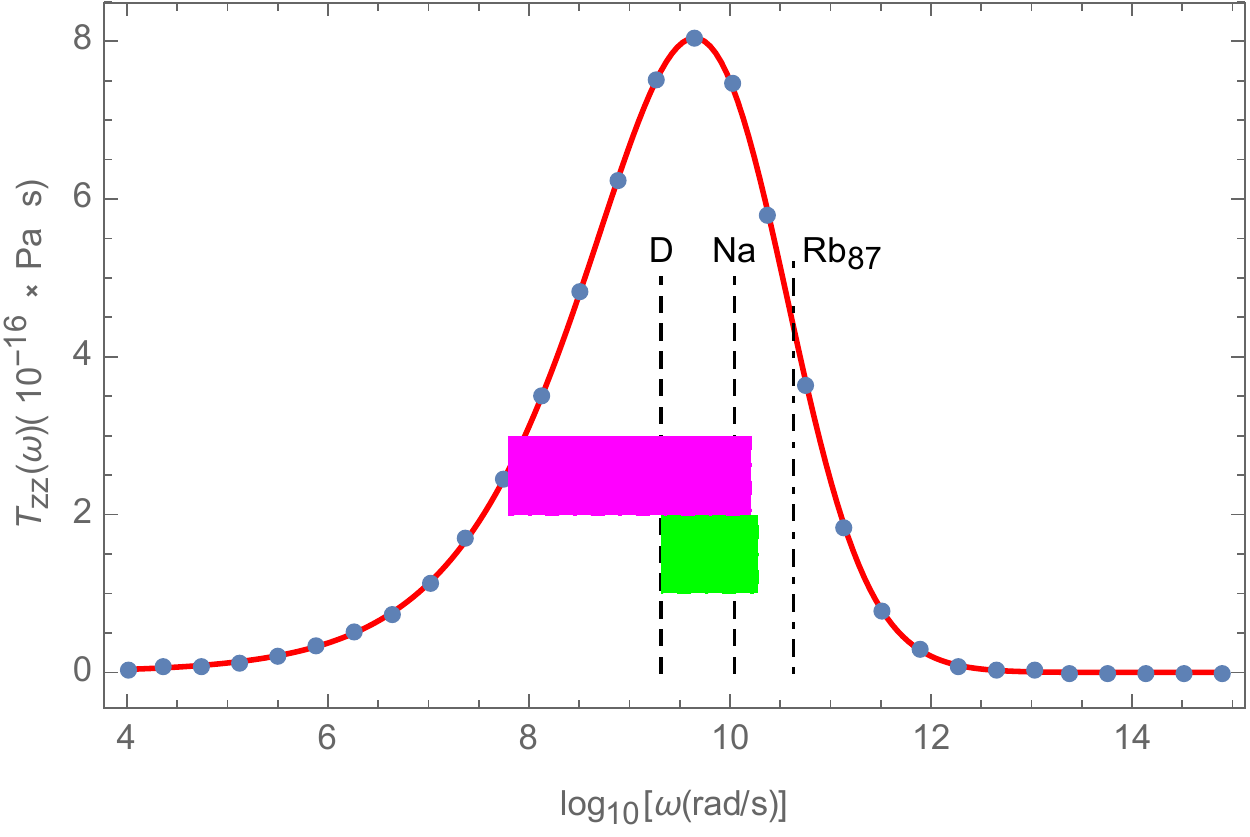}
\caption{\label{spectrum} Spectrum  of the    thermal    Casimir pressure   between two Au thick mirrors at room temperature, described as lossy conductors ($a=2\;\mu$m). The  dots show the contribution from the thermally excited magnetic field.  If the mirrors are modeled as dissipation-less plasmas, the spectrum of the thermal pressure is negligibly small for the displayed frequencies. The dashed vertical lines from left to right correspond to the angular frequencies of the transitions between the ground state hyperfine sub-levels of D, Na, and $^{87}$Rb,  respectively. The upper magenta and the lower green horizontal bands correspond respectively to the angular frequencies of the hyperfine transitions  $\sigma_3$ and   $\sigma_1$  (see Fig. \ref{levels})    between the Zeeman sub-levels of the $1S_{1/2}$ state of D, for values of the external magnetic field in the interval 10 G $< B <$ 1,000 G.}
\end{figure}

The conundrum of the missing repulsive thermal Casimir force  raises the suspicion that the  picture of  thermal  magnetic noise  of a lossy cavity,  composing ${u}^{(\rm cav)}_{\rm TE}$,  is not entirely right.  This prompted us to investigate whether it is possible to  observe  the {\it spectrum}  depicted in Fig. \ref{energy}, to establish which among the lossy Drude or the loss-less plasma  model provides the correct description. 
To provide useful information on the thermal Casimir force, the band-width of the probe    should fall  in the region of frequencies    which contribute significantly to the unobserved thermal Casimir force. To this end, we computed the spectrum  ${\cal T}_{zz}(\omega)$ of the   
thermal Casimir pressure $\Delta P_{\rm C}(a;T)$. The spectrum  ${\cal T}_{zz}(\omega)$ can be obtained starting from the representation of Lishitz formula as an integral  over the real frequency axis \cite{torgerson,BBKM}.  One finds that  for  $a \ll \lambda_T$, ${\cal T}_{zz}(\omega)$ is well described by the formula:
\be
 {\cal T}_{zz} (\omega )= \frac{\hbar}{2 \pi} \;\frac{2\,{\cal H}^{\rm (sc)}_{\perp}(\omega;z) 
-{\cal H}^{\rm (sc)}_{||}(\omega;z) }{ \exp(\hbar \omega/k_B T)-1}  
 \;,\label{specas}
\ee
where we set 
\begin{eqnarray}
{\cal H}^{\rm (sc)}_{\perp}(\omega;z)&\!=\!&{\rm Im}[{\cal H}^{\rm (sc)}_{x x}({\bf r},{\bf r},\omega)]={\rm Im}[{\cal H}^{\rm (sc)}_{y y}({\bf r},{\bf r},\omega)]\;,\nonumber\\
{\cal H}^{\rm (sc)}_{||}(\omega;z)&=&{\rm Im}[{\cal H}^{\rm (sc)}_{z z}({\bf r},{\bf r},\omega)]\;.
\end{eqnarray}
The expressions of the scattering contributions ${\cal H}^{\rm (sc)}_{\alpha \beta}({\bf r},{\bf r},\omega)$ to the magnetic Green functions ${\cal H}_{\alpha \beta}({\bf r},{\bf r},\omega)$ can be found in the Appendix.    Equation (\ref{specas}) shows explicitly the magnetic character of the thermal force.
A plot of  ${\cal T}_{zz}(\omega)$ is shown in Fig.\ref{spectrum}. By comparing Fig.\ref{spectrum} with Fig.\ref{energy}, we see that from the point of view of the Casimir force, the relevant frequencies are those towards the high end of the thermal spectrum, corresponding roughly to angular frequencies in the range from $10^7$ rad/s to $10^{11}$ rad/sec.

\section{Experimental scheme}

The  above considerations suggest  that   atomic {\it hyperfine} transitions  could be used  to probe the thermal spectrum of the cavity. There are three distinct reasons supporting this belief.  First of all, atoms are small and therefore they  do not appreciably perturb the thermal field  of the  cavity.  Second, since  a magnetic field   couples to atomic  magnetic moments, hyperfine transitions are sensitive to magnetic field fluctuations.  Third, and more importantly,   the  frequencies of hyperfine atomic transitions typically belong to the GHz region, which coincides with   the peak of the spectrum of the thermal Casimir force  of a micron-sized cavity   (see Fig. \ref{spectrum}).   We underline that the magnetic trap experiment  \cite{cornell} does not provides much information on the problem of interest for the present work, because the  frequencies  probed by this experiment are in the MHz region, which are two to three orders of magnitude smaller than the  frequencies of interest for the thermal Casimir force.  In addition to that, the experiment  involved a single metallic surface, and therefore it is not clear to what extent its results can be extrapolated to a Casimir system of two closely spaced metallic surfaces.   The possible existence of saturation effects in a cavity  makes it desirable to observe the spectrum of the thermal em field in the gap of a micron-sized metallic cavity, an enterprise that has never been attempted so far.  In principle, one might consider an experimental scheme similar to that used in the experiment Ref.\cite{haroche},  based on the observation of the shift of the hyperfine spectral lines of the  atoms passing between the mirrors,   caused by the Casimir-Polder interaction of the atoms with the fluctuating magnetic  field of the cavity. Unfortunately, estimates of the shifts of hyperfine levels suffered by atoms placed near a metallic surface  \cite{carsten2}
indicate that the effect is too small to be measurable. Thus, a different approach is needed.

\begin{figure}
\includegraphics [width=.9\columnwidth]{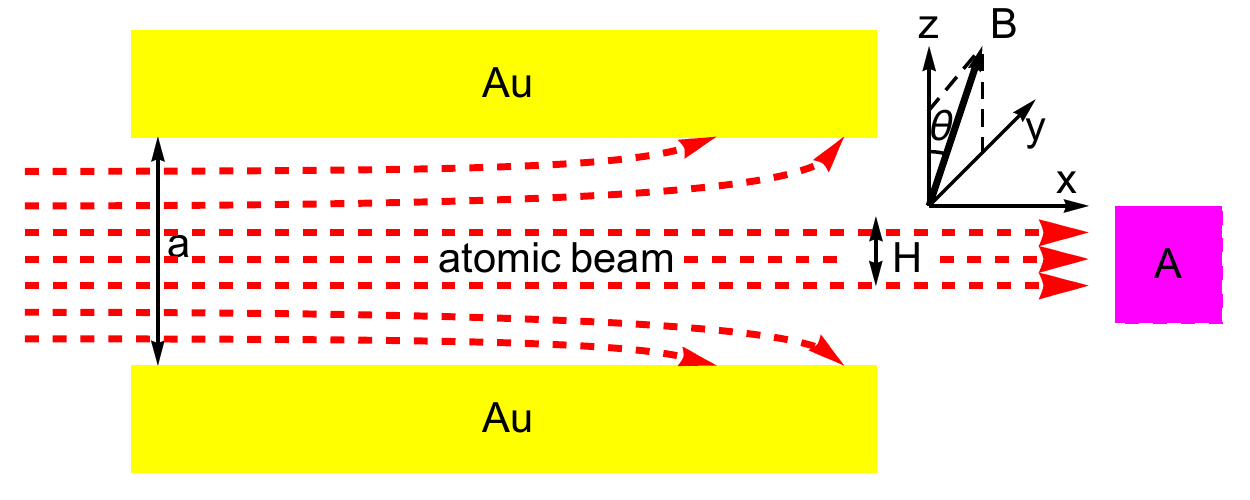}
\caption{\label{setup} Experimental scheme.  A plane parallel Au cavity of width $a=2\;\mu$m is placed in a uniform magnetic field $\bf B$ directed in the $(y,z)$ plane,  forming an angle $\theta$ with the $z$-axis. A  spin-polarized beam of D atoms prepared in a  single Zeeman hyperfine $1S_{1/2} (F,m_F)$ state,    traverses the cavity in the $x$ direction.  Thermally excited magnetic fields inside the cavity cause  transitions   $(F,m_F) \rightarrow (F',m'_F)$ among ground state hyperfine sublevels. Only atoms passing within a narrow  channel of width $H$ across the center can traverse the cavity without being deflected onto the mirrors by the Casimir-Polder interaction. The populations of the hyperfine sublevels  by the atoms that escape the cavity are measured by the detector $A$.}
\end{figure}

The scheme of the experiment we propose is in its general lines similar to that of Ref. \cite{jhe}, and is illustrated in Fig. \ref{setup}: a room-temperature planar cavity of width $a=2\;\mu$m   consisting of two parallel Au mirrors in a high  vacuum   is placed in a uniform magnetic field ${\bf B}$, forming an angle $\theta$ with the $z$ axis.  A monochromatic beam of spin-polarized  D atoms of  velocity $v$, initially prepared in a single Zeeman hyperfine state $1S_{1/2}(F,m_F)$ \footnote{With respect to the quantization axis provided by the magnetic field, the $2(2 I+1)$ non-degenerate Zeeman hyperfine sublevels are labelled, as usual, by the  quantum numbers $(F,m_F)$ of the limiting  zero-field states \cite{ramsey}.}  enters the space between the mirrors, moving in the direction $x$ perpendicular to the plane containing the  ${\bf B}$-field and the $z$ axis. We let $L=$ 1 cm the length of the mirrors in the beam direction.   If the mirrors are modeled as  lossy conductors, the  strong magnetic noise  in the gap between the mirrors causes a significant number of transitions  $(F,m_F) \rightarrow (F',m'_F)$ among ground state hyperfine sublevels. If instead dissipation is neglected, the magnetic noise gets suppressed by many orders of magnitudes, and practically no transitions occur (apart from those caused by collisions of the atoms with the residual gas that is present in the vacuum chamber). A clear cut discrimination among the two models is in principle possible by measuring with the detector $A$ the  populations of the hyperfine levels $(F',m'_F)$ by the D atoms that leave the cavity. Ideally, for the test to have a high confidence level,   the   atoms should spend inside the cavity a sufficient  time for many transitions to occur, under the hypothesis that the Drude model is correct. This goal can be achieved either by taking a longer cavity, or by slowing down the atoms. There is a {\it caveat}, however.  As it was noted already in \cite{jhe,haroche},  the atoms inside the cavity are subjected to the Casimir-Polder (CP) attraction of the mirrors. As a result, the atoms that pass too far from the center, where the CP force vanishes by symmetry, end up colliding with the mirrors, where they stick, and are thus removed from the beam. Of course, this happens more easily  for slow atoms and/or for a long cavity, and so one has to do a compromise.  The best chances of success are offered by heavy atoms with weak CP interaction, a criterion that led us to the choice of D atoms. We estimated that to have a transition probability of say 5  \%, the  D atoms should  traverse the cavity with a velocity of 20 m/s.  Using available data for the polarizability of ground state H atoms in \cite{cebim} to compute the CP potential $U_{\rm CP}(z_0)$  \cite{buh}  of the 1S state of D (up to  a negligible correction, the hyperfine states $(F,m_F)$ have the same CP potential \cite{carsten2}), we found that for a velocity of 20 m/s the only D atoms  that escape from the cavity  are those  passing in a narrow channel of width $H=100$ nm across the center (see Fig.\ref{setup}). 
This means that approximately  5 \% of the  D atoms  escape the cavity without being pulled onto the mirrors by the CP force.

The {\it radiative} \footnote{In our analysis we neglect the contribution $\Gamma_{\rm fi}^{(\rm coll)}$ of atomic collisions, which depends on the residual pressure of the vacuum chamber. The data reported in the experiment \cite{jhe} show that atomic collisions have a small effect on hyperfine transitions in a high vacuum. The rate  $\Gamma_{\rm fi}^{(\rm coll)}$  could in principle be measured  by using  a non-metallic cavity of same geometry, and then subtracted from the data collected with the Au cavity.} transition rate $\Gamma_{\rm fi}$ from the initial state $|i\rangle = (F,m_F)$ to the final state $|f\rangle = (F',m'_F)$ of an atom placed at the position ${\bf r}_0$ in the gap between the mirrors can be computed using the  formalism of \cite{wylie}:
\be
\Gamma_{\rm fi}=\mu_{\alpha}^{\rm i f}\mu_{\beta}^{\rm  fi} \int_{-\infty}^{\infty} dt \,e^{i \omega t} \langle {\hat B}_{\alpha}({\bf r}_0,t) {\hat B}_{\beta}({\bf r}_0,0)\rangle\;,
\label{rate0}
\ee
where $\omega=(E_{\rm i}-E_{\rm f})/\hbar$ is the transition frequency, and $\mu_{\alpha}^{\rm i f}$ is the matrix element of the atomic magnetic dipole moment operator:
\be
{ {\vec \mu}}^{\;\rm  f i}=\langle f |\, \left[ g_{\rm D} \mu_{\rm N}  {\hat {\vec I}} - \mu_{\rm B}( g_{\rm e}{\hat  {\vec S}}+{\hat {\vec L}})\right]\, |i \rangle\;.
\ee
Here, $g_{\rm e}=2.0023$ and  $g_{\rm D}=0.857407$  \cite{ramsey} are the gyromagnetic factors of the electron and of the D nucleus,  respectively, $\mu_{\rm N}$ and $\mu_{\rm B}$ are the nuclear and Bohr magnetons,  ${\hat {\vec I}}$ is the nuclear spin (I=1 for D), and ${\hat  {\vec S}}$ and ${\hat {\vec L}}$ are, respectively, the electron's spin and orbital angular moment (all angular momenta are   in units of $\hbar$).  The correlation functions in Eq. (\ref{rate0}) are not symmetrized, differently from those that appear in Eq. (\ref{corr}). The two types of correlators are  however in a simple relation with each other \cite{agarwal}: 
$$
\langle {\hat B}_{\alpha}({\bf r},t) {\hat B}_{\beta}({\bf r'},t') \rangle=\langle {\hat B}_{\alpha}({\bf r},t) {\hat B}_{\beta}({\bf r'},t') \rangle_{\rm sym}
$$
\be
+ \hbar \int_{-\infty}^{\infty} \frac{d \omega}{2 \pi}  {\rm Im}[{\cal H}_{\alpha \beta}({\bf r},{\bf r'},\omega)]e^{-i \omega(t-t')}\;.
\ee
Substituting the above relation into Eq. (\ref{rate0}), and using Eq. (\ref{corr}),  one obtains the following formula for the rate:
\be
\Gamma_{\rm fi}=\frac{2}{\hbar (1-e^{-\hbar \omega/k_B T})}\mu_{\alpha}^{\rm i f}\mu_{\beta}^{\rm  fi} \; {\rm Im}[{\cal H}_{\alpha \beta}({\bf r}_0,{\bf r}_0,\omega)]\;.\label{rate}
\ee 
From the above formula, one can get the total transition rate  $\Gamma_{i}=\sum_{\rm f \neq i} \Gamma_{\rm fi}$.
Substituting the imaginary part of the free space Green function  $ {\rm Im}[{\cal H}^{(0)}_{\alpha \beta}({\bf r}_0,{\bf r}_0,\omega)  ]= (2/3) (\omega/ c)^3  \delta_{\alpha \beta}$ into the r.h.s. of Eq. (\ref{rate}), one can verify that the  rates $\Gamma^{(\rm free)}_{\rm i}$  outside the cavity are extremely small.  For example, for $T=300$ K, and $B=10$ G the rate of the highest hyperfine state $(F=3/2,m_F=3/2)$  is $\Gamma^{(\rm free)}_{(3/2,3/2)}=$ 1.1$\times 10^{-12}\;{\rm s}^{-1}$. The transition rates inside the gap are obtained  by substituting into the r.h.s. of Eq. (\ref{rate}) the Green functions of the cavity.    By a simple computation, one finds:  
$$
\Gamma^{\rm (cav)}_{\rm fi}= \frac{2}{\hbar (1-e^{-\hbar \omega/k_B T})}
$$
$$
\times \left\{ \left[ {\cal H}^{\rm (cav)}_{\perp}(\omega;z_0)\left(1+\cos^2 \theta \right) + {\cal H}^{\rm (cav)}_{||}(\omega;z_0) \sin^2 \theta  \right] |\mu_{x}^{\rm i f}|^2 \right. 
$$
\be
+ \left. \left[ {\cal H}^{\rm (cav)}_{\perp}(\omega;z_0) \sin^2 \theta   + {\cal H}^{\rm (cav)}_{||}(\omega;z_0) \cos^2 \theta  \right] |\mu_{\zeta}^{\rm i f}|^2 \right\}\;, \label{ratec}
\ee
where  $\mu_{\zeta}^{\rm i f}={\vec \mu}^{\;\rm i f}\cdot {\vec B}/|B|$. The coefficients $ {\cal H}^{\rm (cav)}_{\perp}(\omega;z_0)$ and $ {\cal H}^{\rm (cav)}_{||}(\omega;z_0)$ are defined in a similar manner as $ {\cal H}^{\rm (sc)}_{\perp}(\omega;z_0)$ and $ {\cal H}^{\rm (sc)}_{||}(\omega;z_0)$: 
\begin{eqnarray}
{\cal H}^{\rm (cav)}_{\perp}(\omega;z)&\!=\!&{\rm Im}[{\cal H}^{\rm (cav)}_{x x}({\bf r},{\bf r},\omega)]={\rm Im}[{\cal H}^{\rm (cav)}_{y y}({\bf r},{\bf r},\omega)]\;,\nonumber\\
{\cal H}^{\rm (cav)}_{||}(\omega;z)&=&{\rm Im}[{\cal H}^{\rm (cav)}_{z z}({\bf r},{\bf r},\omega)]\;.
\end{eqnarray}
The two sets of coefficients are related to each other by the following Equations:
\begin{eqnarray}
{\cal H}^{\rm (cav)}_{\perp}(\omega;z)&\!=\!&{\cal H}^{\rm(sc)}_{\perp}(\omega;z)+ 2\, \omega^3 / 3 c^3 \;,\nonumber\\
{\cal H}^{\rm (cav)}_{||}(\omega;z)&=& {\cal H}^{\rm (sc)}_{||}(\omega;z)+ 2\, \omega^3 / 3 c^3 \;.
\end{eqnarray}
We note that the formula for $\Gamma^{\rm (cav)}_{\rm fi}$ involves the same coefficients $ {\cal H}^{\rm (sc)}_{\perp}$ and $ {\cal H}^{\rm (sc)}_{z}$  that enter into $ {\cal T}_{zz} (\omega )$ (see Eq. (\ref{specas})). It is apparent from Eqs. (\ref{ratec}) that by appropriate measurements of  $\Gamma^{\rm (cav)}_{\rm fi}$     it is in principle possible to determine the coefficients  $ {\cal H}^{\rm (cav)}_{\perp}$ and $ {\cal H}^{\rm (cav)}_{||}$,  and then to obtain an estimate of the density of the thermal Casimir pressure ${\cal T}_{zz} (\omega )$.

\begin{figure}
\includegraphics [width=.9\columnwidth]{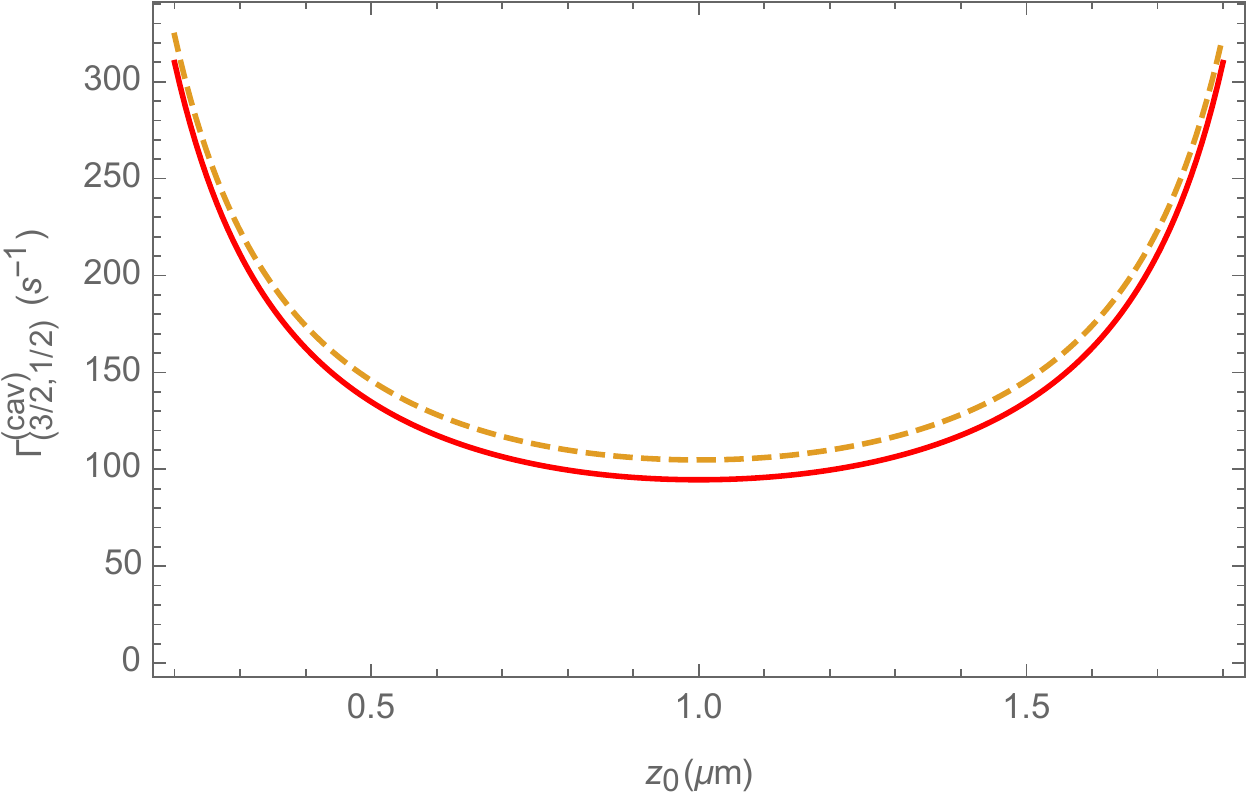}
\caption{\label{ratez}  Radiative rate $\Gamma^{(\rm cav)}_{(3/2,-1/2)}$   in a $2\,\mu$m Au cavity  at $T=300$ K, as a function of the atom coordinate $z_0$. The cavity is placed in a 10 G magnetic field, directed along the $z$ axis (solid line) or along the $y$ axis (dashed line). The Au mirrors are modeled as lossy conductors. If dissipation is neglected, both  rates become smaller than $10^{-12}\;s^{-1}$ for all displayed values of $z_0$.}
\end{figure}

Figure \ref{ratez} shows that the transition rates $\Gamma^{\rm (cav)}_{\rm i}(z_0)$  depend on the atoms's coordinate $z_0$ inside the cavity,  increase rapidly as the atoms approach the mirrors, and diverge at each mirror surface. 
 At first sight,  it would seem that the $z_0$ dependence of the decay rates introduces a large uncertainty in the final  state of the atoms exiting the cavity.  Fortunately, this potential source of uncertainty is actually  negligible, if  one recalls that only atoms passing  in the narrow channel of width $H=$100 nm across  the cavity center  can escape it, without being pulled onto the mirrors by the CP attraction.  
Since the $z_0$-dependence of the rates $\Gamma^{\rm (cav)}_{\rm i}(z_0)$ is flat around the center (see Fig. \ref{ratez}), the  50 nm uncertainty  on the coordinate $z_0$ of the atoms  entails an uncertainty much smaller than one percent in the decay rates $\Gamma^{(\rm cav)}_{\rm fi}$, showing that the atoms that escape the cavity have  almost identical  rates.   
\begin{figure}
\includegraphics [width=.9\columnwidth]{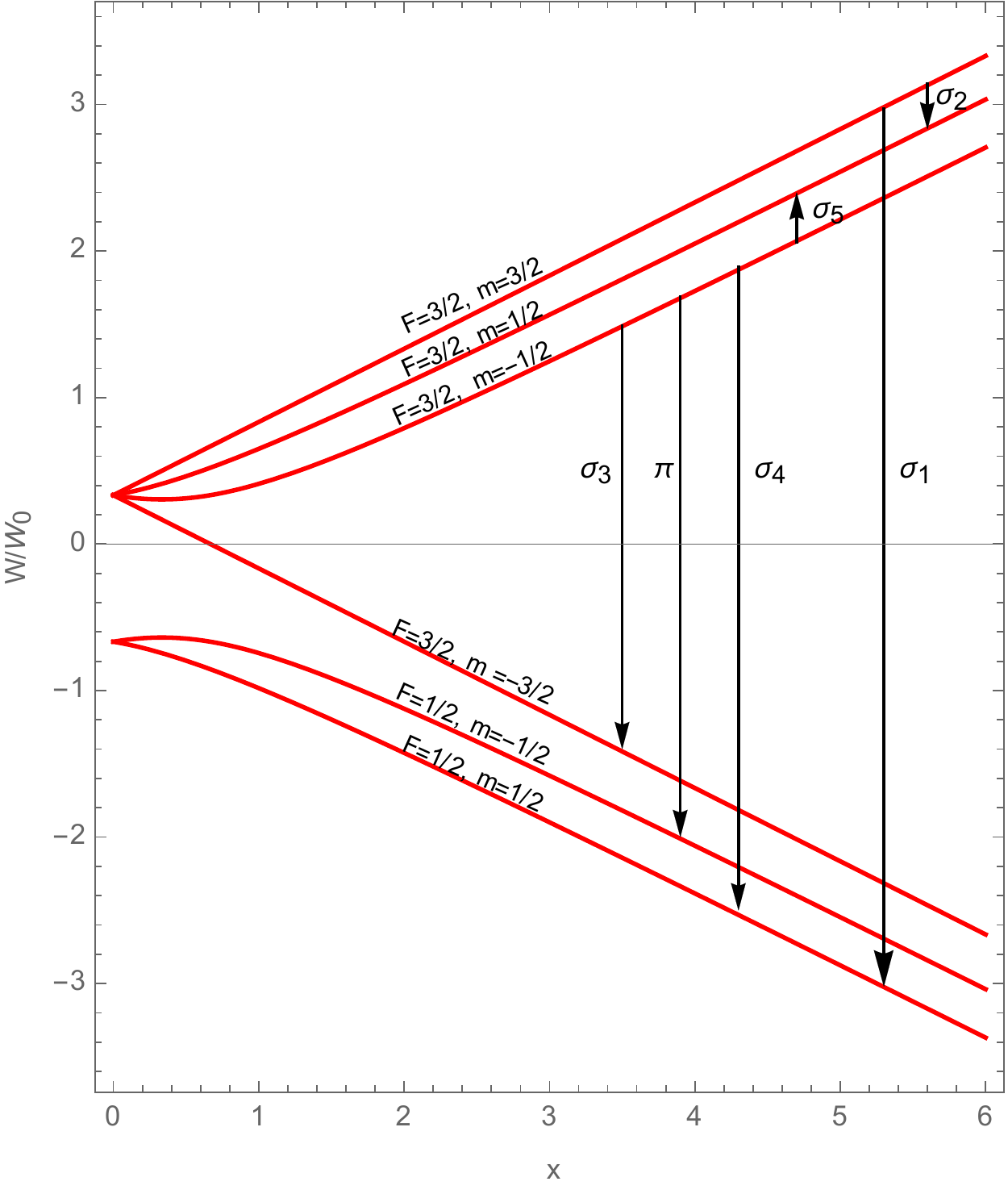}
\caption{\label{levels} Normalized energies of the Zeeman hyperfine sublevels of the $1S_{1/2}$ level of D, in an external magnetic field $B$, versus $x=B\, (g_{\rm N} \,\mu_{\rm N}+g_{\rm e} \,\mu_{\rm B})/W_0$. $W_0=1.354 \times 10^{-6}$ eV is the hyperfine splitting among the $F=3/2$ and the $F=1/2$ sublveles of the  $1S_{1/2}$ state of D in zero field. The vertical arrows show the magnetic-dipole-allowed transitions for the initial states $(F=3/2,m_F=3/2)$, and $(F=3/2,m_F=1/2)$. 
}
\end{figure} 

\begin{figure}
\includegraphics [width=.9\columnwidth]{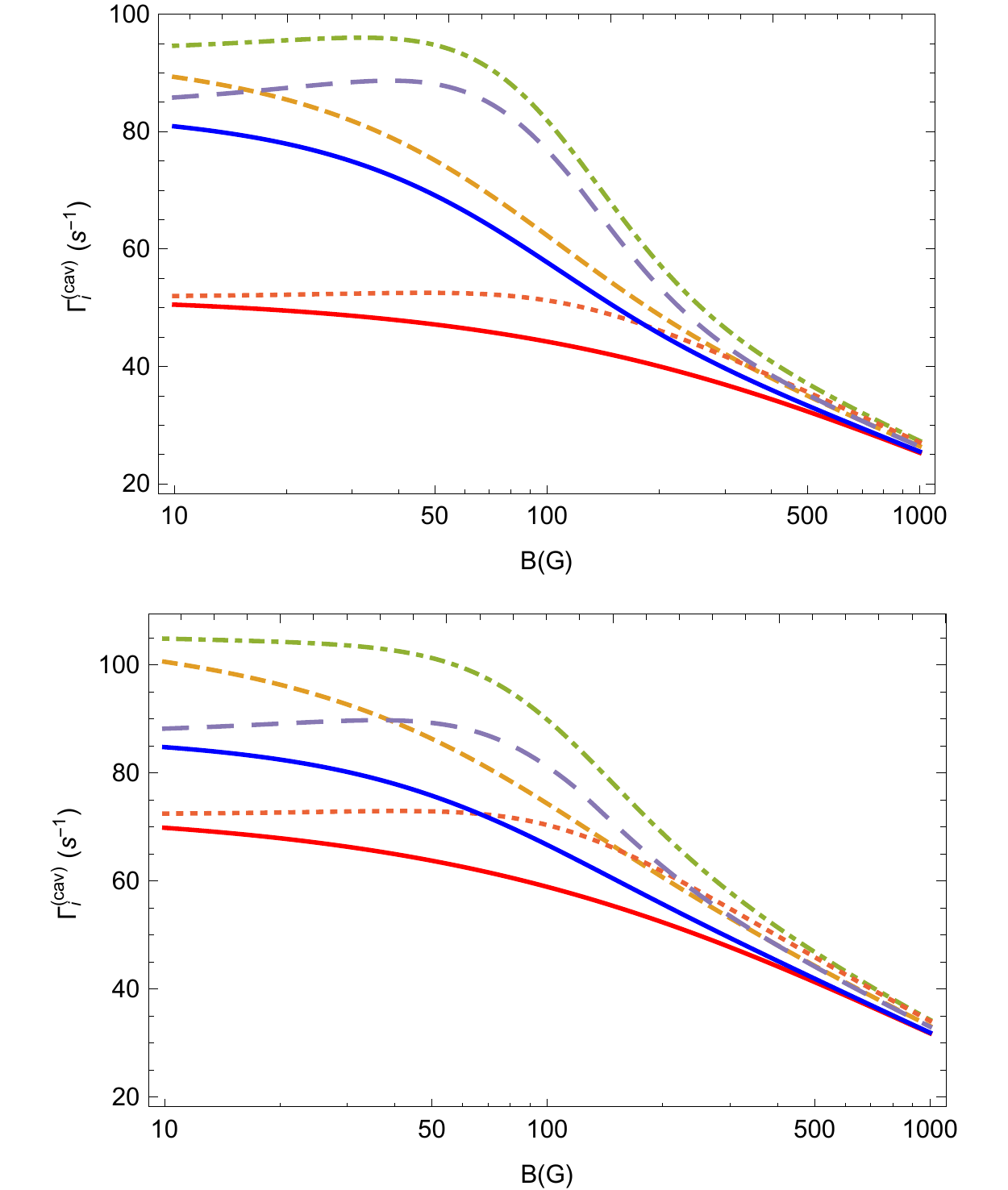}
\caption{\label{rateB}  Radiative rates $\Gamma^{(\rm cav)}_{(F,m_F)}$  at the center of a $2\,\mu$m  Au cavity  for $T=300$ K, as a function of  $B$  (in G).  The Au mirrors are modeled as  lossy conductors. The magnetic field is directed along the $z$-axis (top panel) or along the y-axis (lower panel). The lower solid red, dashed, dotdashed, dotted, long dashed and the upper solid blue lines correspond, respectively, to the Zeeman hyperfine levels of Fig.\ref{levels}, from top to bottom. If the mirrors are modeled as dissipation-less all rates are negligibly small.}
\end{figure}

\begin{figure}
\includegraphics [width=.9\columnwidth]{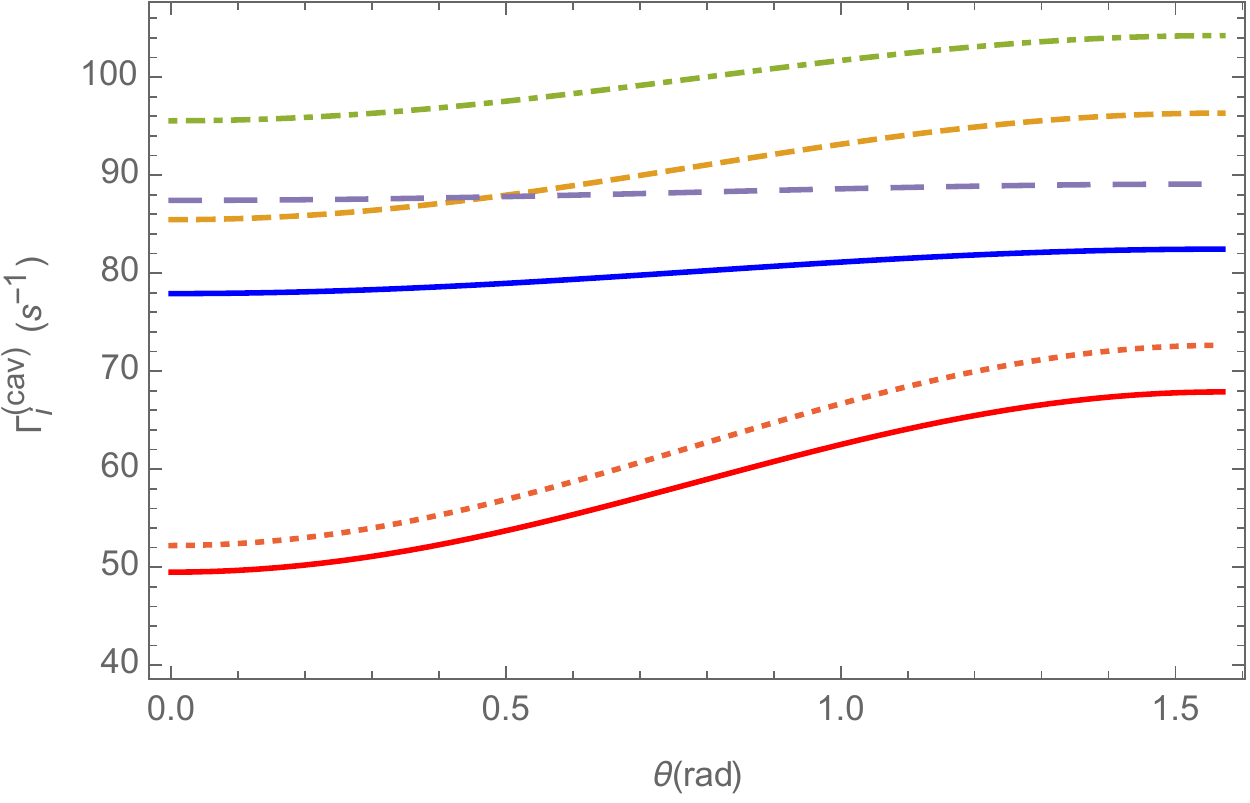}
\caption{\label{ratetheta}  Radiative rates $\Gamma^{(\rm cav)}_{(F,m_F)}$  at the center of a $2\,\mu$m  Au cavity  for $T=300$ K, as a function of  $\theta$  (in rad) for $B=20$ G.  The Au mirrors are modeled as  lossy conductors.   The lower solid red, dashed, dotdashed, dotted, long dashed and the upper solid blue lines correspond, respectively, to the Zeeman hyperfine levels of Fig.\ref{levels}, from top to bottom. If the mirrors are modeled as dissipation-less all rates are negligibly small.}
\end{figure}

The  radiative rates   $\Gamma_{\rm i}^{(\rm cav)}$  depend dramatically on the  used model for the low-frequency electric permittivity $\epsilon(\omega)$ of the Au mirrors. If the mirrors are described by the plasma model, the rates  turn out to be unmeasurably small, and   of similar  magnitudes as those in free space. For example, for   $T=300$ K and with a field $B=10$ G directed along the $y$ axis, one finds $\Gamma^{(\rm cav)}_{(3/2,-1/2)}\vert_{\rm pl}=$ 5$\times 10^{-13}\;{\rm s}^{-1}$.
If the mirrors are modeled as Drude conductors, the transition rates
increase by many orders of magnitude. In Fig. \ref{rateB} we show plots of the Drude-model radiative rates $\Gamma^{\rm (cav)}_{\rm i}$ at the center of the cavity, versus $B$  (in G). The magnetic field is directed along the $z$-axis (top panel) or along the $y$-axis (lower panel). In Fig. \ref{ratetheta} the radiative rates are plotted as a function of the angle $\theta$, for fixed $B=20$ G. In both figures, the solid lower red, dashed, dotdashed, dotted, long dashed and the upper solid blue  lines correspond, respectively, to the Zeeman hyperfine levels of Fig.\ref{levels}, from top to bottom. 
We  see that the Drude-model values of the rates differ by more than thirteen orders of magnitudes from the plasma model ones. Since the Larmor frequencies   can be tuned  by varying the strength of the magnetic field $B$ (see Fig. \ref{levels}),  it is possible in this way to scan  a large part of the frequency interval that contributes to the thermal Casimir force  (see the colored bands in Fig. \ref{spectrum}). Figres \ref{rateB} and \ref{ratetheta}  show that the  hyperfine state with the largest transition rate is  $(F=3/2,m_F=-1/2)$, which suggests to prepare the atoms in this state.

A precise  determination of the probability distribution of the atoms exiting the cavity   can be obtained by solving  the master equation \cite{buhman}, which describes the evolution  of the  probabilities $(p_1(t), \cdots p_{2(2 I+1)}(t))$   to find  an atom in any of the $2(2 I+1)$  hyperfine sublevels, during    the time interval $0 < t <\tau$ it spends   in the cavity:
\be
\frac{d p_{n}}{d t}=- p_n \sum_{k \neq n} \Gamma^{(\rm cav)}_{kn} + \sum_{k \neq n} \Gamma^{(\rm cav)}_{nk} p_k\;.\label{master}
\ee
By solving this Equation one finds for example that for a 10 G field directed along the $y$-axis,   D atoms prepared in the state $(3/2,-1/2)$   decay into other hyperfine states with a total probability of about 5 \%.  The  final probability distribution among the states $(F',m'_F)$ is the following: $p_{(1/2,1/2)}=0.57\%$, $p_{(1/2,-1/2)}=1.07\%$, $p_{(3/2-3/2)}=1.57\%$, $p_{(3/2,-1/2)}=94.9\%$, $p_{(3/2,1/2)}=1.84\%$ and $p_{(3/2,3/2)}= 0.02\%$.

Larger transition probabilities  can be clearly achieved, if necessary, by considering atoms of lower velocity. The price to pay in that case is however that  less atoms shall be able to traverse the cavity, without getting deflected onto the mirrors before reaching the exit. It is not easy to tell a priori what is the best compromise among these competing   factors.

\section{Conclusions}

In conclusion,  we have shown that a micron-sized metallic cavity is filled with  non-resonant radiation having transverse electric polarization, following a non-Planckian spectrum,  whose average density at room temperature  is  orders of magnitudes larger than that of a black-body at the same temperature. Differently from the typical $T^4$  dependence of the energy of a large black-body cavity, the energy density of a  narrow cavity whose width $a$ is smaller than the thermal length $\lambda_T$, displays an almost linear dependence on the temperature. Computations show that the existence of this radiation is strictly  dependent on the dissipative properties of real mirrors,   an that no such radiation exists in an ideal cavity with no losses. The mechanical pressure exerted by this radiation on the mirrors  coincides with the repulsive thermal correction to the Casimir force, predicted by Lifshitz theory for two lossy plates at finite temperature \cite{sernelius}. The actual existence of this thermal force is much debated, since several precision Casimir experiments with metallic surfaces failed to observe it. 

We have shown that the  spectrum of  this radiation can be measured by observing the transition rates  between hyperfine ground-state sublevels $1S_{1/2}(F,m_F) \rightarrow 1S_{1/2}(F',m'_F)$  of D atoms passing in the gap between  the mirrors. Apart from providing a  test of the extension of Planck's law to the sub-wavelength regime, such a  measurement would  shed much light on the puzzle of the missing thermal Casimir force, which remains unsolved after twenty years. 
The availability of  optical techniques which allow  to manipulate and observe with exquisite precision  cold atoms  in well  determined  Zeeman hyperfine states \cite{wieman,Datom,wang} makes us hopeful that the experiment described in this work is feasible with current apparatus.

\acknowledgments

The author thanks T. Emig, M. Kardar, M. Kr\"uger and R. L. Jaffe for  useful discussions.  

 .
\appendix*

\section{Green function of a planar cavity}

In this Appendix we provide the explict formulae for the Green functions of a cavity, that are needed for the computations described in the present work.

At  points $\bf r$ and ${\bf r}'$ in the gap between two parallel dielectric slabs at distance $a$  in vacuum,  the electric Green function ${\cal E}_{\alpha \beta}({\bf r},{\bf r'},\omega)$  can be decomposed as:
\be
{\cal E}^{(\rm cav)}_{\alpha \beta}({\bf r},{\bf r'},\omega)={\cal E}^{(0)}_{\alpha \beta}({\bf r},{\bf r'},\omega)+{\cal E}^{(\rm sc)}_{\alpha \beta}({\bf r},{\bf r'},\omega)\;, \label{Greensplit}
\ee
where ${\cal E}^{(0)}_{\alpha \beta}({\bf r},{\bf r'},\omega)$ is the free-space Green function, and ${\cal E}^{(\rm sc)}_{\alpha \beta}({\bf r},{\bf r'},\omega)$ is a scattering contribution. The magnetic Green function has an analogous decomposition:
\be
{\cal H}^{(\rm cav)}_{\alpha \beta}({\bf r},{\bf r'},\omega)={\cal H}^{(0)}_{\alpha \beta}({\bf r},{\bf r'},\omega)+{\cal H}^{(\rm sc)}_{\alpha \beta}({\bf r},{\bf r'},\omega)\;. \label{GreensplitH}
\ee
The free-space Green function has the expression:
$$
{\bf \cal E}^{(0)}({\bf r},{\bf r}',\omega)={\bf \cal H}^{(0)}({\bf r},{\bf r}',\omega)=\left[(3 {\hat {\bf R}} \otimes {\hat {\bf R}} - {\bf 1}) \left( \frac{1}{R^3}- \frac{i\,\omega}{c R^2}\right) \right.
$$ 
\be
+\left. ({\bf 1}- {\hat {\bf R}} \otimes {\hat {\bf R}}) \frac{\omega^2}{c^2 R} - \frac{4 \pi}{3} \delta({\bf R}){\bf 1}\right]e^{i \omega R/c}\;,
\ee
where ${\bf R}= {\bf r}-{\bf r}'$. Note that the imaginary part of the free-space Green function is non-singular for ${\bf r} \rightarrow {\bf r}'$:
\be
\lim_{{\bf r} \rightarrow {\bf r}'}{\rm Im}[{\bf \cal E}^{(0)}({\bf r},{\bf r}',\omega)]=\lim_{{\bf r} \rightarrow {\bf r}'}{\rm Im}[{\bf \cal H}^{(0)}({\bf r},{\bf r}',\omega)]=\frac{2 \omega^3}{3 c^3}{\bf 1}\;.
\ee
In the limit   ${\bf r} \rightarrow {\bf r}'$, the scattering part of the Green tensor ${\cal E}^{(\rm sc)}_{\alpha \beta}({\bf r},{\bf r}',\omega)$ attains a finite limit, and its non vanishing components are:
$$
{\cal E}^{(\rm sc)}_{x x}({\bf r},{\bf r},\omega)={\cal E}^{(\rm sc)}_{y y }({\bf r},{\bf r},\omega)=4 \pi i \int \frac{d^2 {\bf k}_{\perp}}{(2 \pi)^2} k_z 
$$
$$
\times \left[\left(\frac{R_{\rm p}^{(1)}  R_{\rm p}^{(2)}}{{\cal A}_{\rm p}}+ \frac{\omega^2}{c^2 k_z^2}\frac{R_{\rm s}^{(1)}  R_{\rm s}^{(2)}}{{\cal A}_{\rm s}}  \right) e^{2 i k_z a} \right. + \frac{1}{2}\left( \frac{\omega^2}{c^2 k_z^2} \frac{R_{\rm s}^{(1)} }{{\cal A}_{\rm s}} \right.
$$
\be
\left.\left. - \frac{R_{\rm p}^{(1)} }{{\cal A}_{\rm p}}\right)e^{2 i k_z z}+ 
 \frac{1}{2}\left( \frac{\omega^2}{c^2 k_z^2} \frac{R_{\rm s}^{(2)} }{{\cal A}_{\rm s}} - \frac{R_{\rm p}^{(2)} }{{\cal A}_{\rm p}}\right)e^{2 i k_z (a-z)}\right]\;, \label{Exx}
\ee
and
$$
{\cal E}^{(\rm sc)}_{z z}({\bf r},{\bf r},\omega)= 4 \pi i \int \frac{d^2 {\bf k}_{\perp}}{(2 \pi)^2} \frac{k_{\perp}^2}{k_z} \left( \frac{R_{\rm p}^{(1)}  R_{\rm p}^{(2)}}{{\cal A}_{\rm p}} e^{2 i k_z a} \right.
$$
\be
\left. + \frac{R_{\rm p}^{(1)} }{2 {\cal A}_{\rm p}}\, e^{2 i k_z z} +  \frac{R_{\rm p}^{(2)} }{2 {\cal A}_{\rm p}} \,e^{2 i k_z (a-z)} \right)\;,\label{Ezz}
\ee
where ${\bf k}_{\perp}$ is the in-plane wave-vector, $k_z=\sqrt{\omega^2/c^2-k_{\perp}^2}$,  the indices $\rm s$ and $\rm p$ denote TE and TM polarizations, respectively, $R_{\alpha}^{(k)}$  is the reflection coefficient of the $k$-th mirror for polarization $\alpha={\rm s}, {\rm p}$ and  ${\cal A}_{\alpha}=1-R_{\alpha}^{(1)} R_{\alpha}^{(2)} e^{2 i k_z a}$. The corresponding formulae for the magnetic Green tensor  ${\cal H}^{(\rm sc)}_{\alpha \beta}({\bf r},{\bf r},\omega)$ can be obtained from those of the electric Green tensor, by interchanging the reflection coefficients  $R_{\rm s}^{(k)} \leftrightarrow R_{\rm p}^{(k)}$ into Eqs. (\ref{Exx}) and (\ref{Ezz}).


\begin{thebibliography}{99}

\bibitem{Planck} M. Planck, Verh. Deutsch. Phys. Ges. {\bf 2}, 202 (1900).

\bibitem{rytov} S.M.\ Rytov, {\it Theory of Electrical Fluctuations and Thermal Radiation}, Publishing House, Academy os Sciences, USSR (1953).

\bibitem{kruger} M. Kr\"uger, G. Bimonte, T. Emig and M. Kardar, Phys. Rev. B {\bf 86}, 115423 (2012). 

\bibitem{mehran}  G. Bimonte, T. Emig, M. Kardar, and M. Kr\"uger, Ann. Rev. Cond. Matt. Phys. {\bf 8}, 119 (2017).

\bibitem{agarwal} G. S. Agarwal, Phys. Rev. A {\bf 11}, 230 (1975).

\bibitem{wylie} J. M. Wylie and J. E. Sipe, Phys. Rev. A {\bf 30}, 1185 (1984).


\bibitem{lifs} E. M. Lifshitz, Zh. Eksp. Teor. Fiz. {\bf 29}, 94 (1955) [Sov. Phys. JETP {\bf 2}, 73 (1956)].

\bibitem{Casimir48} H.~B.~G. Casimir, Proc. K. Ned. Akad. Wet., {\bf 51},  793 (1948).

\bibitem{haroche} V. Sandoghdar, C. I. Sukenik, E. A. Hinds, and S. Haroche, Phys. Rev. Lett. {\bf 68}, 3432 (1992).

\bibitem{buh} S. Y. Buhmann,  \textit{Dispersion Forces I: Macroscopic Quantum
Electrodynamics and Ground-State Casimir, Casimir-Polder, and van der Waals Forces} (Springer, Berlin, 2012).

\bibitem{kleppner} S. Haroche and D. Kleppner, Phys. Today {\bf 42}, 24 (1989).

\bibitem{jhe} W. Jhe,  A. Anderson, E. A. Hinds, D. Meschede, L. Moi, and S. Haroche,
Phys. Rev. Lett. {\bf 58}, 666 (1987).

\bibitem{carsten1} C. Henkel, S. P\"otting, and M. Wilkens, Appl. Phys. B {\bf 69}, 379 (1999). 

\bibitem{carsten3} C. Henkel, P. Kr\"uger, R. Folman, and J. Schmiedmayer, Appl. Phys. B {\bf 76}, 173 (2003).

\bibitem{cornell} D. M. Harber,  J. M. McGuirk, J. M. Obrecht,  and E. A. Cornell, J. Low Temp. Phys. {\bf 133}, 229 (2003).

\bibitem{sernelius} M. B\"ostrom and B.E. Sernelius, Phys. Rev. Lett. {\bf 84}, 4757 (2000).

\bibitem{decca1} R.~S.~Decca,  E.~Fischbach, G.~L.~Klimchitskaya, D.~E.~Krause,
D.~L\'opez, and V.~M.~Mostepanenko, Phys. Rev. D {\bf 68}, 116003 (2003).

\bibitem{decca2} R.{\ }S. Decca, D. L\'opez, E. Fischbach, G.{\ }L. Klimchitskaya,
D.{\ }E. Krause, and V.{\ }M.\ Mostepanenko, Ann. Phys. (N.Y.) {\bf 318}, 37 (2005).

\bibitem{decca3} R.~S.~Decca, D.~L\'opez, E.~Fischbach, G.~L.~Klimchitskaya, D.~E.~Krause, and V.~M.~Mostepanenko, Phys. Rev. D {\bf 75}, 077101 (2007).

\bibitem{decca4} R.~S.~Decca, D.~L\'opez, E.~Fischbach, G.~L.~Klimchitskaya, D.~E.~Krause, and V.~M.~Mostepanenko, Eur. Phys. J. C {\bf 51}, 963 (2007).

\bibitem{chang} C.-C.~Chang, A.~A.~Banishev, R.~Castillo-Garza, G.~L.~Klimchitskaya, V.\ M.\ Mostepanenko, and U.\ Mohideen, Phys. Rev. B {\bf 85}, 165443 (2012).

\bibitem{bani1} A.~A.~Banishev, G.~L.~Klimchitskaya, V.\ M.\ Mostepanenko, and U.\ Mohideen,
Phys. Rev. Lett. {\bf 110}, 137401 (2013).

\bibitem{bani2} A.~A.~Banishev, G.~L.~Klimchitskaya, V.\ M.\ Mostepanenko, and U.\ Mohideen,
Phys. Rev. B {\bf 88}, 155410 (2013).

\bibitem{book2} M. Bordag, G. L. Klimchitskaya, U. Mohideen and V. M. Mostepanenko,   \textit{Advances in the Casimir Effect} (Oxford University Press,  2009).

\bibitem{sernelius2} B. E. Sernelius, Phys. Rev. A {\bf 80}, 043828 (2009).

\bibitem{lamorth}
 A.~O.~Sushkov, W.~J.~Kim, D.\ A.\ R.\ Dalvit, and S.\ K.\ Lamoreaux, Nature Phys. {\bf 7}, 230 (2011).
 
 \bibitem{antonini} P. Antonini, G. Bimonte, G. Bressi, G. Carugno, G. Galeazzi, G. Messineo, and G. Ruoso, J. Phys. Conf. Ser. {\bf 161}, 012006 (2009).

\bibitem{rednik} G.~L.~Klimchitskaya, V.\ M.\ Mostepanenko, R. I. P. Sedmik, and H. Abele, Symmetry {\bf 11}, 407 (2019).

 

\bibitem{landau} L. D. Landau and E. M. Lifshitz, {\it Statistical Physics}, Part 2, Pergamon Press, England (1980).











\bibitem{BimonteJ} G.~Bimonte, New J. Phys. {\bf 9}, 281 (2007).

\bibitem{carsten} F.~Intravaia and C.~Henkel, Phys. Rev. Lett. {\bf 103}, 130405 (2009).

 \bibitem{torgerson}  J. R. Torgerson and S. K. Lamoreaux, Phys. Rev. E {\bf 70}, 047102 (2004).

\bibitem{BBKM} V. B. Bezerra, G. Bimonte, G.~L.~Klimchitskaya, V.\ M.\ Mostepanenko, and C. Romero, Eur. Phys. J. C {\bf 52}, 701 (2007).
 






















 

\bibitem{carsten2} H. Haakh, F. Intravaia, C. Henkel, S. Spagnolo, R Passante, B. Power, and F. Sols, Phys. Rev. A {\bf 80}, 062905 (2009).


\bibitem{ramsey} N. F. Ramsey,  \textit{Molecular Beams} (Oxford Clarendon Press, UK, 1956).


\bibitem{cebim} M. A. Cebim, M. Masimi and J. J. De Groote, Few-Body Syst. {\bf 46}, 75 (2009).



\bibitem{buhman} S. Y. Buhmann and S. Scheel, Phys. Rev. Lett. {\bf 100}, 253201 (2008).

\bibitem{wieman} B. P. Masterson, C. Tanner, H. Patrick, and C. E. Wieman, Phys. Rev. A {\bf 47}, 2139 (1993).

\bibitem{Datom} D. Szczerba, L.D. van Buuren, J. F. J van den Brand, H. J Bulten, M. Ferro-Luzzi, S. Klous. H. Kolster, J. Lang, F. Mul, H. R. Poolman, and M. C. Simani, Nucl. Instr. and Meth.  A {\bf 455}, 769 (2000).

\bibitem{wang} B. Wang, Y. Han, J. Xiao, X. Yang, C. Zhang, H. Wang, M. Xiao, and K. Peng, Phys. Rev. A {\bf 75}, 051801(R) (2007).
 

\end{thebibliography}
\end{document}